\documentclass[]{aa}  
\usepackage{graphicx,psfig}
\usepackage{txfonts}
\begin{document}
   \title{A search for transiting extrasolar planet candidates \\in the OGLE-II microlens database of the galactic plane}

   \author{Snellen I.A.G.\inst{1}, van der Burg R.F.J. \inst{1}, de Hoon M.D.J.\inst{1}, Vuijsje F.N.\inst{1}}

   \offprints{snellen@strw.leidenuniv.nl}

   \institute{Leiden Observatory, Leiden University, Postbus 9513, 2300 RA, Leiden, The Netherlands}

   \date{}

 
  \abstract{In the late 1990s, the Optical Gravitational Lensing Experiment (OGLE) team 
conducted the second phase of their long-term monitoring programme, OGLE-II, which 
since has been superseded by OGLE-III. All the monitoring data of this second phase, 
which was primarily aimed at finding microlensing events, have recently been made 
public.} 
{Fields in the OGLE-II survey have typically been observed with a cadence of 
once per night, over a period of a few months per year. We investigated whether these 
radically differently sampled data can also be used to search for transiting extrasolar planets,
in particular in the light of future projects such as PanSTARRS and SkyMapper, which will
monitor large fields, but mostly not at a cadence typical for transit studies.}
{We selected data for $\sim$15700 stars with 13.0$<$I$<$16.0 in three OGLE-II 
fields towards the galactic disc in the constellation Carina, each with 500-600 epochs of 
I-band photometry. These light curves were first detrended using Sys-Rem, after which they were 
searched for low amplitude transits using the Box Least Squares algorithm.}
{The detrending algorithm significantly decreased the scatter in the light curves, 
from an average of 0.5\% down to 0.2$-$0.3\% for stars with $I$$<$15. 
Several dozens of 
eclipsing binaries and low amplitude transits were found, of which 13 candidates
exhibit transits with such depth and duration  that they are possibly caused
by an object with a radius less than twice that of Jupiter. Eleven out of these 
thirteen candidates show significant ellipsoidal light variations and are 
unlikely to host a transiting extrasolar planet. However, OGLE2-TR-L9 (CAR\_SC2\_75679), 
is an excellent planet candidate comparable to the known OGLE-III transiting planets,
and deserves further follow-up observations.
}
{}


   \maketitle
%

\section{Introduction}

An extrasolar planet that transits its host star is of great scientific value. 
Analysis of its transit light curve can reveal the planet/star size ratio,
and implicitly the radius of the planet after the star has been spectrally typed. 
Since the occurrence of a transit implies that the inclination of the planetary
orbit is near 90$^\circ$, radial velocity measurements will reveal the true 
mass of the planet, in contrast to only a lower limit in mass for non-transiting planets.
This will yield not only the mean density of the planet, but can also give 
insight into its internal structure (eg. Sato et al. 2005). 
In addition, many interesting follow-up opportunities are within reach, such as 
atmospheric transmission spectroscopy, secondary eclipse and Rossiter-McLaughlin effect
observations, and transit timing, providing information on the planet's atmosphere,
orientation of its orbit, and on the presence of other planets in the system
(eg. Charbonneau et al. 2002; Charbonneau et al. 2005; Deming et al. 2005; Winn et al. 2007; 
Snellen 2004; Steffen et al. 2007).

Several successful transit surveys are now in place. 
OGLE-III (the third phase of the Optical Gravitational
Lensing Experiment; Udalski et al 2002a,b,c;2004) is the longest running survey and 
has yielded so far 5 transiting planets. While more than 150 planet candidates have
been reported by this team, most of them turned out to be blended or grazing eclipsing 
binaries and/or eclipsing M-dwarfs (Bouchy et al. 2005; Pont et al. 2005). It shows that radial velocity follow-up is 
crucial for the identification of the genuine transiting extrasolar planets.
 Now, several other dedicated ground-based 
transit surveys, such as XO (McCullough et al. 2005), SuperWASP (Collier Cameron et al. 2007), 
TrES (O'Donovan et al. 2007), and HAT (Bakos et al. 2007), have
found their first planets.  Therefore, the number of confirmed 
transiting planets has recently increased significantly (20 in total), with more than half of 
them discovered in the last 1.5 years. However, the number of planets found to transit the 
stars being monitored still
lags behind the expectations from radial velocity surveys (Gaudi 2006). 
In addition, almost all of the planets found so far have 
radii larger than that of Jupiter. This indicates that a significant fraction of the transiting hot 
Jupiters are being missed. The most likely cause has been identified as
correlated noise in the light curve data sets (e.g. Pont et al. 2006; Aigrain \& Pont 2007). 
Variations in atmospheric (seeing, airmass) or instrumental/telescope
properties result in photometric features on the time scale of a transit, causing 
the sensitivity of a survey to decrease significantly. The recently launched CoRoT
satellite and future Kepler mission will likely resolve these issues.

In this paper we present a search for transiting planet candidates in the 
1997$-$2000 OGLE-II survey (Udalski et al. 1997), a precursor of OGLE-III, which was not 
designed
for transit searches. While it is in itself highly desirable to find more 
transiting exoplanets, it is also valuable in the light of the radically 
different  time sampling of OGLE-II compared to dedicated transit surveys 
(typically once per night compared to once per 15 minutes), which should 
significantly reduce correlated noise problems. This is particularly interesting
in the light of future projects such as PanSTARRS (e.g. Afonso \& Hennig 2007) and 
SkyMapper (eg. Bayliss \& Sackett 2007); groundbased telescopes with huge 
field of view that will monitor millions of stars, but mostly at a cadence 
very different from transit surveys (although they will probably also spend some
limited time performing transit surveys).
In the next section the OGLE-II survey and our data analysis are presented.
In section 3 the results are shown and discussed, and our conclusions are given in 
section 4. The folded light curves are presented in Figure \ref{curves} in the appendix.

\begin{figure}
\psfig{figure=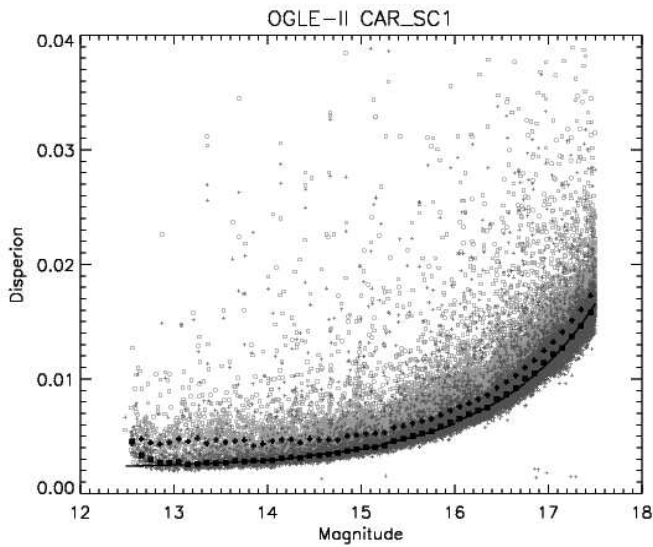,width=0.5\textwidth}
\caption{\label{dispersion} The dispersion in the stellar light curves as function of the median magnitude before (light grey points) and after (dark grey points) the use of Sys-Rem. The black points indicate the median dispersion per 0.1 magnitude bin
before and after this detrending correction. It shows that the algorithm is most efficient for the brightest stars. For the bright stars (I$<$15) the photometric uncertainties are decreased by a factor of $\sim$2.
 }
\end{figure}

\section{OGLE-II survey and data analysis}

OGLE-II was the precursor of the current OGLE-III project, dedicated to 
microlensing work. All of its photometric data has been made available by
the OGLE team on http://ogledb.astrouw.edu.pl/$\sim$ogle/photdb/ 
(Szymanski 2005). The data have been
collected with the 1.3m Warsaw Telescope at the Las Campanas observatory in 
Chile, using a 2048x2048 CCD camera in drift scan mode. Details on the 
instrumental setup can be found in Udalski et al. (1997).
The observed fields cover 14.2$\times$57 arcmin in the sky. 
All images were reduced by the OGLE team using the standard OGLE pipeline
(Udalski et al. 1998). 
During the OGLE-II phase, fields have been observed towards the galactic bulge,
the LMC and SMC, and the galactic disc. In this paper, we concentrate 
our analysis on the three galactic disc fields in the constellation 
Carina, which are the only disc fields which have been observed for more 
than 400 epochs, all taken between January 1997 and July 2000.
The relevant field parameters are given in Table \ref{fields}.

\begin{table}
\caption{\label{fields} Details on the OGLE-II fields used in our analysis,
with in column 1 the OGLE-II name, in column 2 the field center, 
column 3 the number of stars 13.0$<$I$<$16.0, and in column 4 the number of 
epochs}
\begin{tabular}{lcccc}
 Name & Ra  & Dec (J2000) & N$_{\rm{stars}}$&N$_{\rm{epochs}}$\\
      &    h  m    & d  m        &                 &\\
CAR\_SC1 &11 06 & -61 30  &5531&502\\
CAR\_SC2 &11 08 & -61 30  &5265&562\\
CAR\_SC3 &11 10 & -61 00  &4915&485\\
\end{tabular}
\end{table}

\begin{table*}
\caption{\label{table2} Candidate low luminosity and planetary transits in the OGLE2 Carina fields towards the galactic disc.
}
\begin{tabular}{llcccrccccccccccc}\\
Candidate & OGLE-II Name & RA(J2000) & Dec (J2000) & I & S$_{\rm{T}}$/$\sigma$ & P & T$_0$\\
          &              & h m s & d m s & & mag & days &  (HJD)     \\ 
OGLE2-TR-L1  &CAR\_SC1\_59908  &11 05 51.70&  $-$61 32 48.7&14.780&11.8  &2.7801006$\pm$1.5e-5& 479.009$\pm$0.0014\\
OGLE2-TR-L2  &CAR\_SC1\_66174  &11 05 56.28&  $-$61 27 35.3&15.647&9.5   &0.6619433$\pm$1.3e-6& 477.077$\pm$0.0003\\
OGLE2-TR-L3  &CAR\_SC1\_121984 &11 06 10.75&  $-$61 14 52.9&15.919&16.8  &1.5841842$\pm$3.0e-6& 478.074$\pm$0.0008\\
OGLE2-TR-L4  &CAR\_SC1\_138478 &11 06 50.39&  $-$61 51 50.3&14.831&10.3  &2.0867921$\pm$8.0e-6& 476.687$\pm$0.0010\\
OGLE2-TR-L5  &CAR\_SC1\_144353 &11 06 33.74&  $-$61 43 17.0&15.980&23.5  &1.8214055$\pm$4.0e-6& 476.818$\pm$0.0009 \\
OGLE2-TR-L6  &CAR\_SC1\_167415 &11 06 57.49&  $-$61 17 19.8&15.519&9.7   &0.8177890$\pm$9.0e-6& 476.864$\pm$0.0004\\
OGLE2-TR-L7  &CAR\_SC1\_169832 &11 06 51.21&  $-$61 11 10.3&15.340&8.1   &1.8038728$\pm$7.0e-6& 477.777$\pm$0.0009\\
OGLE2-TR-L8  &CAR\_SC2\_71230  &11 07 46.96&  $-$61 18 05.6&14.993&25.1  &0.8525568$\pm$1.7e-6& 477.040$\pm$0.0004\\
OGLE2-TR-L9  &CAR\_SC2\_75679  &11 07 55.29&  $-$61 08 46.3&13.974&15.5  &2.4855300$\pm$1.7e-5& 478.661$\pm$0.0012\\
OGLE2-TR-L10 &CAR\_SC2\_78119  &11 07 53.46&  $-$61 04 18.2&15.890&21.1  &0.8602108$\pm$1.7e-6& 477.150$\pm$0.0004\\
OGLE2-TR-L11 &CAR\_SC2\_145417 &11 08 37.25&  $-$61 20 16.6&14.627&16.0  &2.3088322$\pm$7.1e-6& 478.076$\pm$0.0011\\
OGLE2-TR-L12 &CAR\_SC2\_152898 &11 08 49.96&  $-$61 10 51.1&15.582&12.3  &3.8237386$\pm$4.2e-5& 479.461$\pm$0.0019\\
OGLE2-TR-L13 &CAR\_SC3\_155530 &11 10 35.90&  $-$60 39 49.1&14.848&18.9  &0.8836109$\pm$1.0e-6& 476.663$\pm$0.0004 \\
\\
\end{tabular}

In column 1 the candidate name is given, in column 2 the OGLE-II name, in columns 3 and 4 the RA and Dec (J2000), in column 5 the mean $I$ band magnitude, in column 6 the signal-to-noise ratio of the transit, and in column 7 and 8 transit period and the time of transit (-2450000d) respectively, with their 1$\sigma$ uncertainty.
\end{table*}

\begin{table*}
\caption{\label{table3} The results of least-squared fitting of the folded light curves. }
\begin{tabular}{llcccrclll}\\
Candidate& Depth & D/P & R$_{\rm{1min}}$ & R$_{\rm{2min}}$ & E$_{1}$ & E$_{2}$ & J & H & K\\
           &\% & &    R$_{\rm{Sun}}$ & R$_{\rm{Jup}}$ &  $10^{-3}$ &  $10^{-3}$& \\ 
OGLE2-TR-L1 &1.2$\pm$0.1&0.038$\pm$0.003&1.0&1.1&$+$0.2$\pm$0.3&$-$1.1$\pm$0.3&14.009$\pm$0.026&13.727$\pm$0.022&13.606$\pm$0.044\\
OGLE2-TR-L2 &0.6$\pm$0.1&0.119$\pm$0.005&1.3&1.0&$-$2.2$\pm$0.5&$-$0.3$\pm$0.4&14.175$\pm$0.041&13.688$\pm$0.060&13.267$\pm$0.044\\
OGLE2-TR-L3 &3.4$\pm$0.1&0.062$\pm$0.003&1.2&2.2&$+$0.2$\pm$0.8&$-$1.5$\pm$0.8&14.367$\pm$0.032&13.954$\pm$0.058&13.582$\pm$0.054\\
OGLE2-TR-L4 &0.9$\pm$0.1&0.056$\pm$0.003&1.3&1.3&$-$2.3$\pm$0.3&$+$0.3$\pm$0.3&14.051$\pm$0.029&13.788$\pm$0.044&13.548$\pm$0.051\\
OGLE2-TR-L5 &3.2$\pm$0.1&0.049$\pm$0.002&0.9&1.7&$-$0.0$\pm$0.5&$-$2.0$\pm$0.5&15.094$\pm$0.057&14.754$\pm$0.068&14.546$\pm$0.119 \\
OGLE2-TR-L6 &0.7$\pm$0.1&0.102$\pm$0.003&1.3&1.1&$-$2.4$\pm$0.5&$-$0.9$\pm$0.4&12.849$\pm$0.026&11.537$\pm$0.022&11.035$\pm$0.019\\
OGLE2-TR-L7 &0.6$\pm$0.1&0.065$\pm$0.003&1.4&1.1&$-$0.7$\pm$0.4&$-$1.1$\pm$0.4&14.641$\pm$0.065&14.308$\pm$0.081&14.206$\pm$0.080\\
OGLE2-TR-L8 &1.3$\pm$0.1&0.122$\pm$0.004&1.7&2.0&$-$2.8$\pm$0.4&$-$0.7$\pm$0.3&13.382$\pm$0.026&12.538$\pm$0.027&12.252$\pm$0.029\\
OGLE2-TR-L9 &1.1$\pm$0.1&0.038$\pm$0.004&0.9&0.9&$-$0.2$\pm$0.3&$+$0.3$\pm$0.2&13.508$\pm$0.032&13.246$\pm$0.035&13.117$\pm$0.037\\
OGLE2-TR-L10&2.7$\pm$0.1&0.077$\pm$0.003&0.9&1.5&$-$2.0$\pm$0.5&$+$0.0$\pm$0.5&15.078$\pm$0.054&14.746$\pm$0.081&14.602$\pm$0.119\\
OGLE2-TR-L11&1.1$\pm$0.1&0.060$\pm$0.002&1.6&1.7&$-$0.9$\pm$0.3&$-$1.5$\pm$0.2&14.160$\pm$0.063&14.066$\pm$0.082&13.809$\pm$0.074\\
OGLE2-TR-L12&1.6$\pm$0.1&0.045$\pm$0.005&1.8&2.2&$-$1.8$\pm$0.5&$-$0.4$\pm$0.4&14.707$\pm$0.071&14.391$\pm$0.089&13.829$^{\dagger}$\\
OGLE2-TR-L13&1.1$\pm$0.1&0.073$\pm$0.003&0.8&0.9&$-$2.0$\pm$0.3&$+$0.5$\pm$0.3&14.127$\pm$0.079&13.726$\pm$0.136&13.670$\pm$0.137\\
\multicolumn{4}{l}{$^{\dagger}$ No 2MASS error given due to blend.}\\
\\
\end{tabular}

In column 1 the candidate name is given, in column 2 the depth of the transit, in column 3 the length of the transit, in column 4 and 5 the fitted minimum radius of the primary and secondary object, assuming a zero impact parameter for the transit and assuming that the primary is on the main sequence. Columns 6 and 7 give the fitted amplitude of possible ellipsoidal light variation, with periods half and equal to the orbital period 
respectively. Columns 8, 9, and 10 show the 2MASS J, H, and K magnitudes.
\end{table*}

For our analysis we used the photometric data obtained
from the Difference Image Analysis (DIA; Alard \& Lupton 1998; Alard 2000).
We concentrated on those stars that have good data for more than 98\%
of the epochs, and downloaded all stars with a mean magnitude 
of $I<17.5$ (52703 stars in the three fields). In a later stage, 
our transit analysis only focuses on those stars with $13.0<I<16.0$ 
(15711 objects). 

\begin{figure}
\psfig{figure=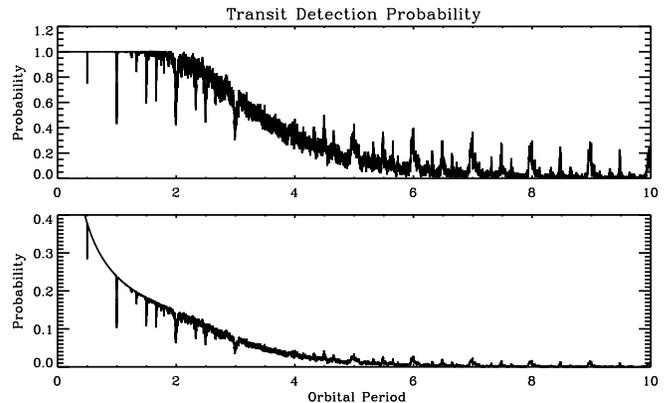,width=0.5\textwidth}
\caption{The transit detection probability for the OGLE-II data as a function
of transit period, assuming a Jupiter-size planet transiting a solar-type star of I=16.0. In the lower panel this probability is multiplied by 
the probability that a planet actually transits. \label{probability}}
\end{figure}

Sys-Rem, a detrending algorithm designed to remove systematic effects 
in a large set of light curves (Tamuz, Mazeh \& Zucker 2005), 
was used on the data from the three fields. The algorithm can detect any 
systematic effect that appears linearly in many light curves obtained by the survey,
without any prior knowledge of the origin of the effects.
It has become a standard tool in transit survey light curve processing
(eg. Bakos et al. 2006; Collier Cameron et al. 2006). In Figure 
\ref{dispersion} the scatter in the data is shown as a function of the 
mean stellar magnitude, for all stars in the CAR\_SC1 field. 
The light-grey symbols indicate the dispersion before the use of Sys-Rem, 
with the black crosses showing the average in each 0.1 magnitude bin.
The dark grey symbols indicate the same but after 10 cycles of Sys-Rem, 
with the black squares the resulting binned average dispersion. 
The algorithm is more efficient for brighter stars (for which the photometric 
errors are more dominated by systematic effects). For the brightest stars ($I<15$)
the mean dispersion has decreased from 0.5\% before Sys-Rem to 0.2-0.3\% after
the use of Sys-Rem, showing the strength of this algorithm. Several of the 
transits presented in section 3 would not have been found without the use 
of Sys-Rem. The final mean error 
as a function of magnitude is fitted with a three parameter function indicated
by the solid black line, containing a residual noise term (0.22\%), and 
Poisson noise terms from the star and the background sky.
The few brightest stars seem to suffer from non-linearity or saturation 
effects, resulting in an increase of the mean dispersion towards the brightest 
$I<13$ objects.

After ten cycles of detrending, a box-fitting algorithm was used to search the light curves 
for periodic transits,
based on the Box Least Squares (BLS) method (Kovacs et al. 2002). Transit periods between
0.1 and 10.0 days were searched for using a detection threshold of 8$\sigma$. 
An estimate of the sensitivity to transiting hot Jupiters as a function 
of their orbital period is shown in Figure \ref{probability}, with in the upper panel 
the probability to detect a transiting object, and in the lower panel this same
probability multiplied by the chance a planet in such an orbit will actually transit its 
star (assuming a random orbital orientation). For this calculation we assumed only 
Jupiter-size objects orbiting  solar type (I=16.0, $\sigma_I$=0.005) stars, thus requiring 16 in-transit data 
points to reach an 8 $\sigma$ detection.
Note that, in contrast to dedicated transit surveys, these 16 data points will most 
likely come from 16 individual transits, since the time in between observations is 
typically much larger (1 day) than the transit duration (a few hours). 
This also means that correlated noise is much less of a problem, with the notable 
exception for near-integer periods. In the latter case, all data points of 
a certain period in time will fall within one or a few areas in phase space. 
Low level stellar variability or slowly varying instrumental instabilities could,
if one is not careful, easily lead to spurious transit detections. 

Several dozens of eclipsing binaries and low amplitude transiting objects
were found among the $\sim$27000 stars. These transits
 were least square fitted using the equations of 
 Mandel \& Agol (2002), keeping the orbital inclination fixed at 
$i$=90$^\circ$. In this way the degeneracy between stellar radius and 
the transit impact parameter was avoided. By assuming that the primary star
is on the main sequence, with M$_1$/M$_{\rm{sun}}$=R$_1$/R$_{\rm{sun}}$, this 
fit subsequently points to lower limits of both the radii of the transited and transiting
objects. In addition, a linear limb darkening profile was assumed with a coefficient of 
$\mu$=0.5. Those systems that were found to have a transiting object with a minimum radius 
of R$_2$$<$2.5R$_{\rm{jup}}$ remain transiting planet candidates.

We subsequently searched for ellipsoidal light variations in those remaining planet 
candidates, since these fluctuations will point to stellar companions:
if the secondary object is sufficiently massive and in a sufficiently close-in 
orbit, tidal effects will deform the primary star. The change in observed solid 
angle and gravity darkening effects then results in sinusoidal light variations
with a period of half the orbital period (Drake 2003; Sirko \& Paczynski 2003).
Sinusoidal light variations equal to the transit period are also possible and 
searched for, either because of possible heating (reflection) effects, or because the 
actual orbital period is twice as large as assumed, due to the fact that both the 
primary and secondary eclipses are detected.  
For all remaining candidates we also retrieved the J, H, and K near-infrared 
magnitudes from the 2MASS survey (Skrutskie et al. 2006).

\section{Results and Discussion}

Our transiting planet candidates are presented in Tables \ref{table2} and 
\ref{table3}, and in Figure \ref{curves} in the appendix.
Table \ref{table2} gives general information and ephemeris of the objects, 
and Table \ref{table3} presents the outcome of 
the least square fits to the transits and of possible out-of-transit ellipsoidal light 
variations. 

We found a total of 13 interesting transiting objects; 7, 5, and 1 in fields 
SC1, SC2, and SC3 respectively. All show a transit shape consistent with a 
secondary object of possibly planetary radius. Only one object, OGLE2-TR-L9 
(CAR\_SC2\_75679) shows no sign of out-of-transit ellipsoidal light variations
and is our best candidate (see below). Seven out of thirteen objects show 
clear ($>$4$\sigma$) sinusoidal variations indicative of massive companions. 
Three others, OGLE2-TR-L7, OGLE2-TR-L12, and OGLE-TR-L1 show hints of these
variations at 2.8$\sigma$, 3.6$\sigma$, and 3.7$\sigma$ respectively, and 
their transiting objects are therefore also unlikely to be planets. 

One object, OGLE-TR-L3, is also of particular interest (see Figure \ref{curves}). 
It shows a significant secondary eclipse, only identifiable because it is 
not at half time between transits, but at phase -0.37. This must be due
to a non-circular orbit of the transiting object.
The timing of the secondary eclipse implies that the 
orbital eccentricity is e$>$0.2. This is quite remarkable
since the orbital period is only P=1.58 days, implying a short tidal
circulization time scale. This means that it should be a recently 
formed binary, or alternatively, that a third body is present in this system.

\subsection*{OGLE2-TR-L9: An excellent exoplanet candidate}

The transit and ellipsoidal light curve analysis shows that 
OGLE2-TR-L9, with an orbital period of P=2.4855 days,
is our best exoplanet candidate. The least square fit to 
its transit yields a lower limit to the radius of the primary of 0.9 R$_{\rm{sun}}$,
and to the radius of the secondary of 0.9 R$_{\rm{jup}}$.
The optical to near-infrared colours of the star, I$-$J=0.466 and J$-$K=0.391, 
are that of a mid-G star, and therefore consistent with the 0.9 R$_{\rm{sun}}$ 
radius of the primary. We also performed a least square fit using the 
equations of Mandel \& Agol (2002), leaving the impact parameter, $\kappa$, as a free 
variable. If the transit was at a high impact parameter (making the radii of 
the primary and secondary more uncertain), it would form a more 'V' shaped
transit, with the ingress and egress times being relatively large with respect to the 
flatter mid-transit part. In contrast, the transit is best fitted with a low impact 
parameter
of $\kappa$$<0.7$ ($\kappa$=$0.6^{+0.1}_{-0.6}$, R$_1$=1.22$^{+0.19}_{-0.41}$ and 
R$_2$=0.122$^{+0.032}_{-0.041}$), resulting in a planet radius of  $<$1.5R$_{\rm{jup}}$.
Hence, this analysis
makes it very likely we are indeed dealing with a planet size object. 

To assess the probability that OGLE2-TR-L9 is of planetary nature,
we performed the same light curve analysis on all the 117 OGLE-III low amplitude
transit candidates from the 2001 and 2002 campaigns as published in Udalski et al. 
(2002a;c). No ellipsoidal light variations could be found at $>$0.2\% amplitude
level for 74 out of 117 targets. Of these, only 9 are best fitted with an 
transiting object with radius $<$1.5R$_{\rm{jup}}$, of which four turned out 
to be planets. Therefore, with the additional information from the 2MASS photometry 
we estimate the likelihood that OGLE2-TR-L9 is indeed a transiting extrasolar 
planet at $\sim$50\%, and deserves follow-up observations. 

The accuracy in the orbital period is such that, now, seven years after the 
last OGLE-II observations (some 1000 periods later), the uncertainty in the transit 
timing has increased to about 30 minutes. First the transit should be searched 
for and the transit shape measured at 30-40$\sigma$ precision with a 1-2 m telescope.
If the transit shape at high signal-to-noise, and spectral stellar typing is 
still consistent with a transit of planetary nature, radial velocity observations
of this star (I=13.97) should be performed to measure the mass of the secondary.
Plans for these follow-up observations are in place.

\begin{figure}
\psfig{figure=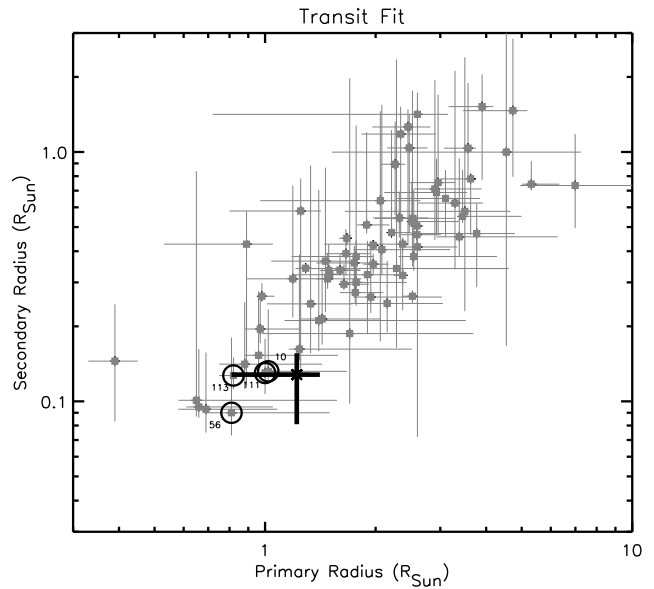,width=0.5\textwidth}
\caption{The radii of the transited and transiting objects from the OGLE-III 2001 
and 2002 campaigns, as estimated from least square fitting of the transit shape 
using the equations of Mandel \& Agol (2002). Only those objects without significant
ellipsoidal light variations are shown. The circles indicate the four confirmed
planets in these samples. The thick black cross indicates the same for the 
planetary candidate, OGLE2-TR-L9, presented in this paper.}
\end{figure}

\section{Conclusions}

In this paper we searched the online database of the second phase  of the OGLE 
project (OGLE-II; 1997-2000) for transiting extrasolar planets, 
concentrating on the galactic disc fields in
the Carina constellation. The main goals were, next to finding possibly more transiting exoplanets, 
to investigate whether such a data set with a very different cadence compared to typical 
transit surveys, is actually sensitive to low amplitude transits. 

We found that detrending the data using the Sys-Rem algorithm significantly reduced the 
scatter in the light curves by about a factor two, down to 0.2\%. About 27000 stars were searched
for transits with periods between 0.5 and 5 days using the BLS algorithm, and several dozens of 
eclipsing binaries were found. Thirteen stars were found to exhibit low amplitude transits,
of which most show significant ellipsoidal light variability indicative of tidal interactions 
and stellar mass secondaries. One star, OGLE2-TR-L9 (P=2.4855 days), shows a transit shape 
consistent with a Jupiter size object orbiting a solar type star, also consistent with its
optical to near-infrared colours. Comparing this candidate to those from the dedicated OGLE-III transit
survey campaigns, we estimate a $\sim$50\% probability for this to be a genuine transiting planet, deserving
follow-up observations.

\section*{Acknowledgements}

We thank the OGLE-team, in particular M. Szymanski, for making the OGLE-II database available online.
This publication also makes use of data products from the Two Micron All Sky Survey, 
which is a joint project of the University of Massachusetts and the Infrared Processing and 
Analysis Center/California Institute of Technology, funded by the National Aeronautics and Space 
Administration and the National Science Foundation.
The Digitized Sky Survey was produced at the Space Telescope Science Institute under U.S. Government 
grant NAG W-2166. The images of these surveys are based on photographic data obtained using the Oschin Schmidt Telescope on Palomar Mountain and the UK Schmidt Telescope. The plates were processed into the present compressed digital form with the permission of these institutions.

\section*{Appendix A}

In this appendix we present the folded light-curves of the thirteen low amplitude 
transiting candidates. The left panel shows for each star the complete light-curve, 
while the middle panel zooms in on $-0.15$$<$phase$<$0.15. The right panel 
is for each star an extract from the Digitized Sky Survey, measuring 2$\times$2 
arcminutes.

\begin{figure*}
\psfig{figure=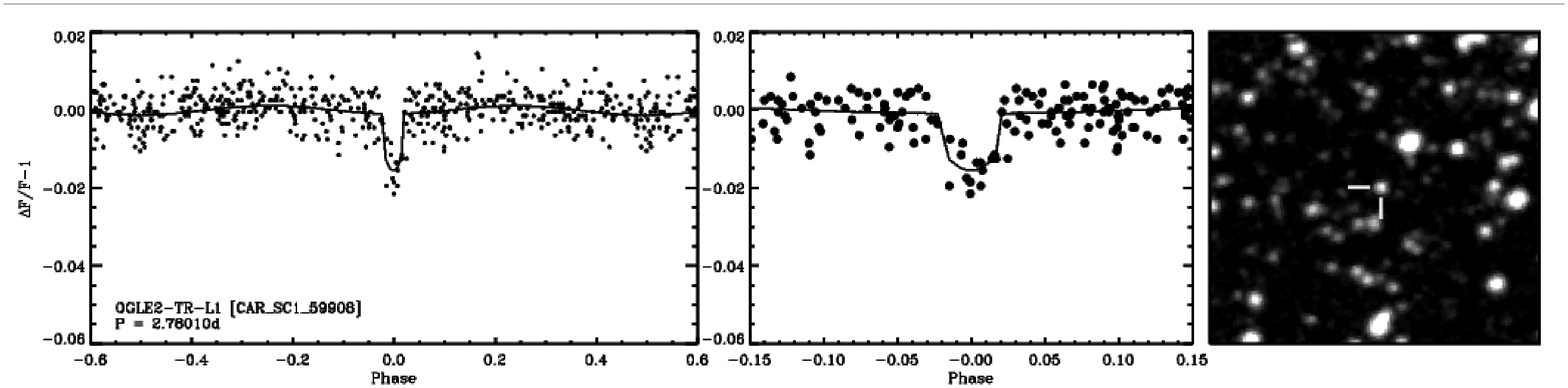,width=1.0\textwidth}
\psfig{figure=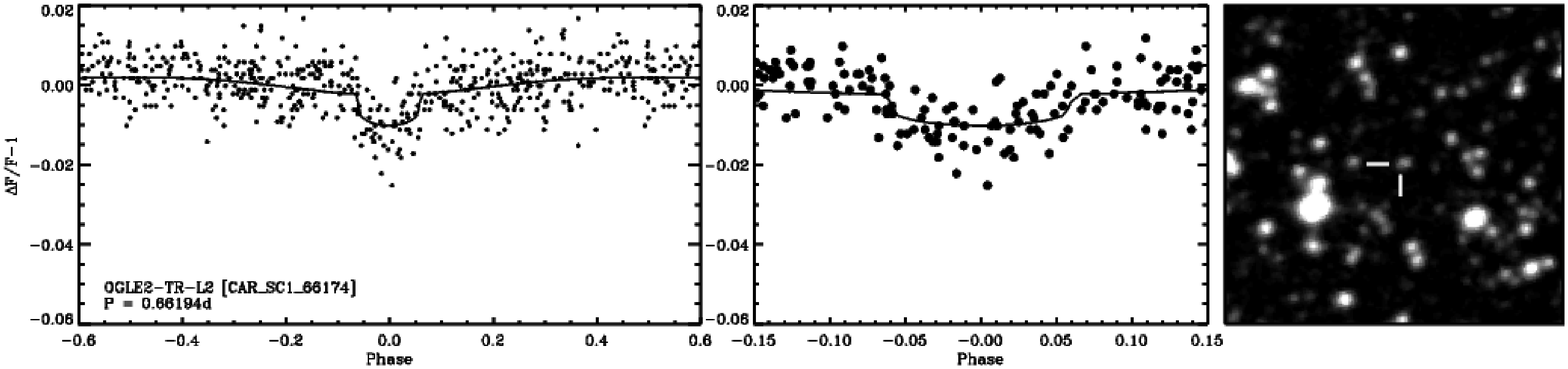,width=1.0\textwidth}
\psfig{figure=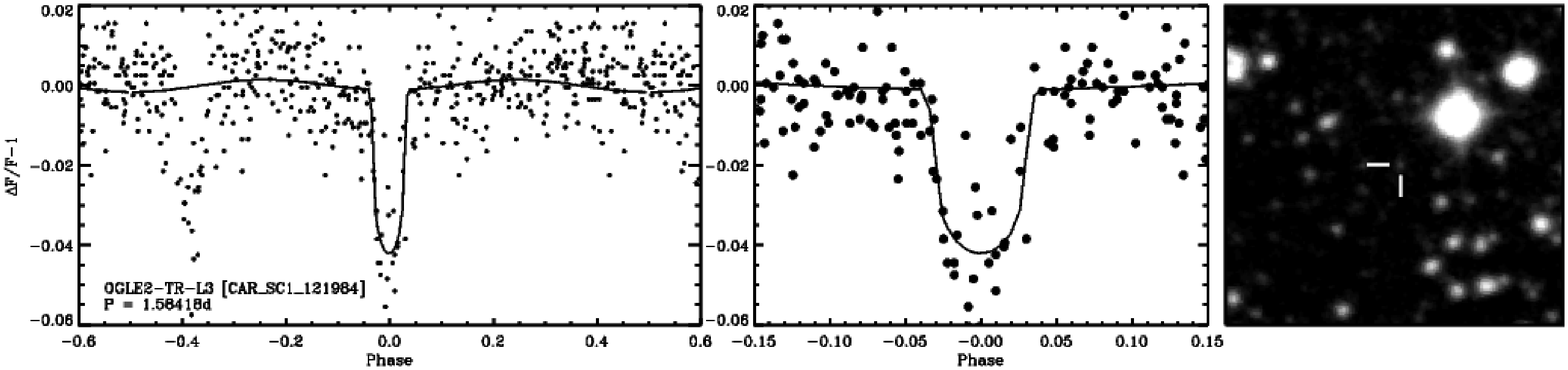,width=1.0\textwidth}
\psfig{figure=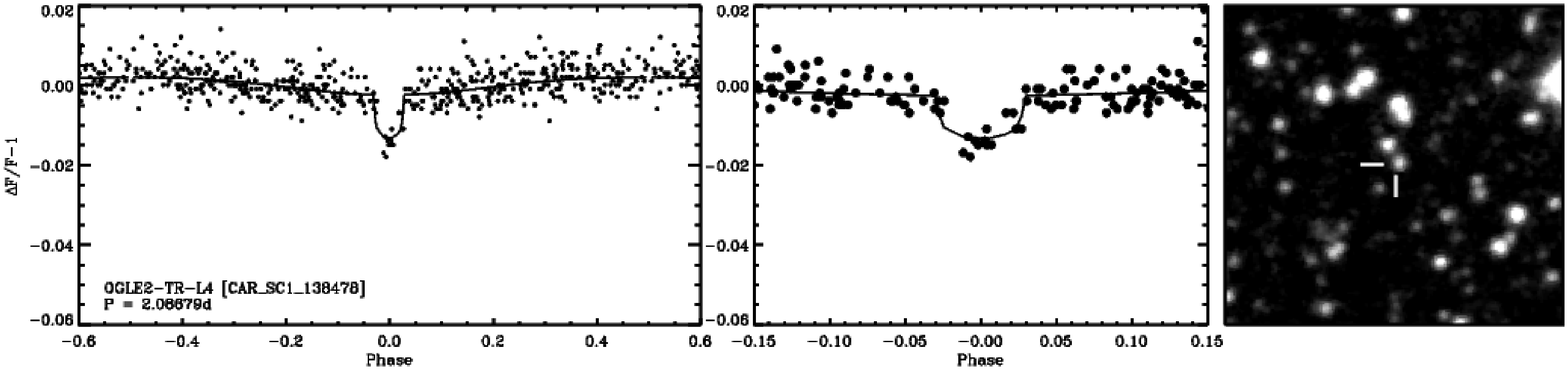,width=1.0\textwidth}
\psfig{figure=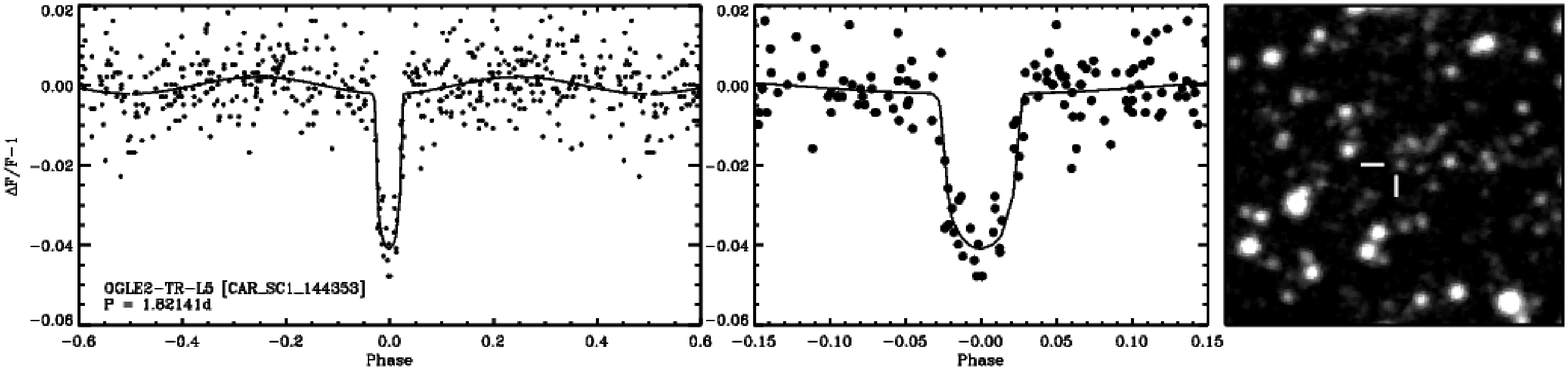,width=1.0\textwidth}
\caption{The folded light curves and finding charts of our transit candidates \label{curves}}
\end{figure*}
\addtocounter{figure}{-1}
\begin{figure*}
\psfig{figure=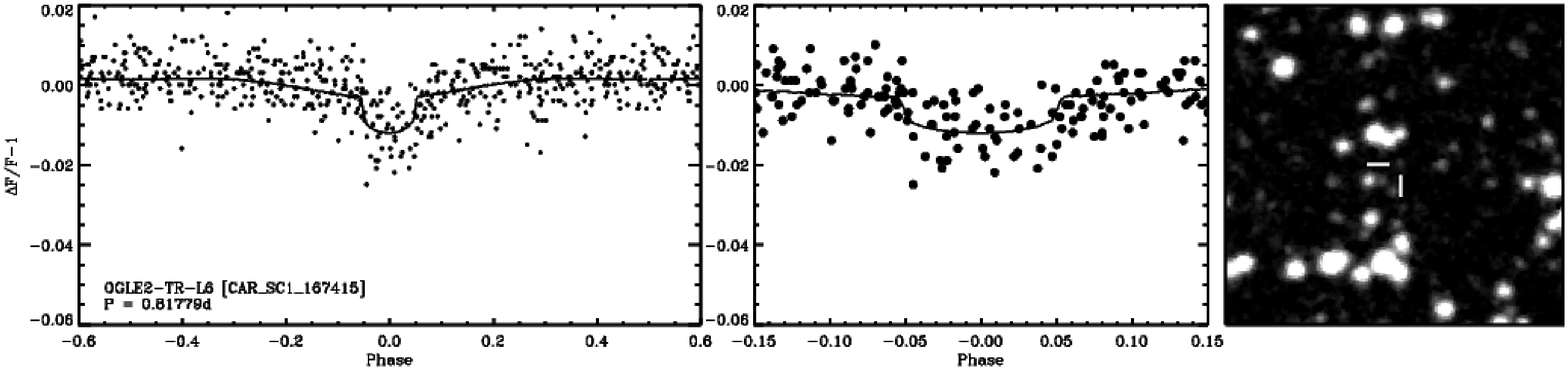,width=1.0\textwidth}
\psfig{figure=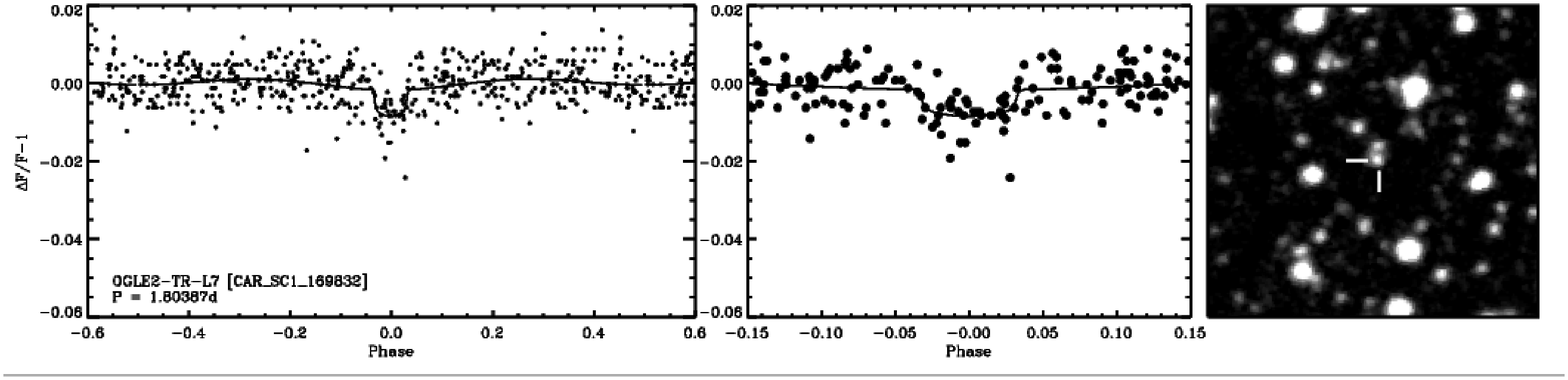,width=1.0\textwidth}
\psfig{figure=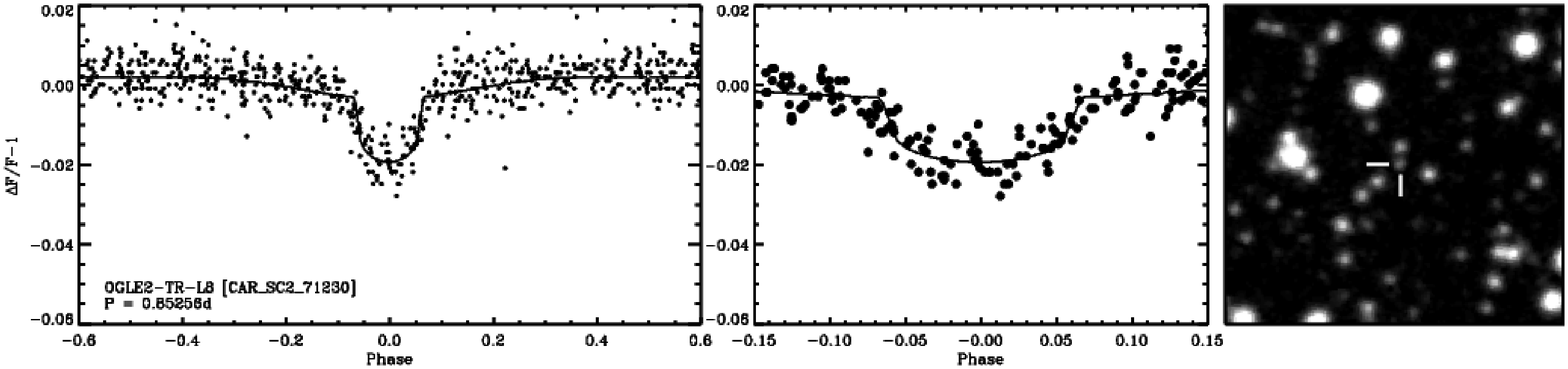,width=1.0\textwidth}
\psfig{figure=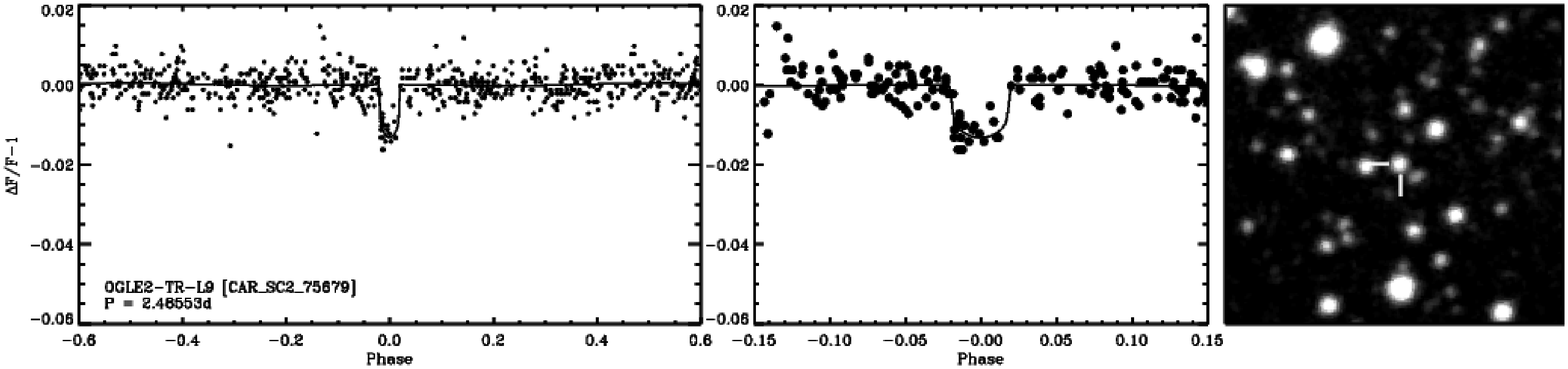,width=1.0\textwidth}
\psfig{figure=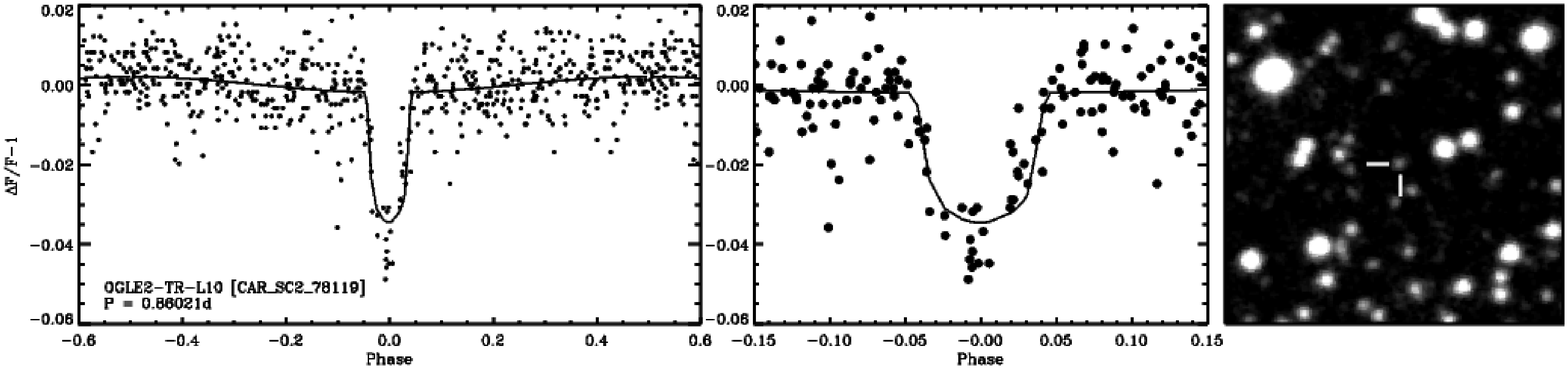,width=1.0\textwidth}
\caption{Continued....}
\end{figure*}
\addtocounter{figure}{-1}
\begin{figure*}
\psfig{figure=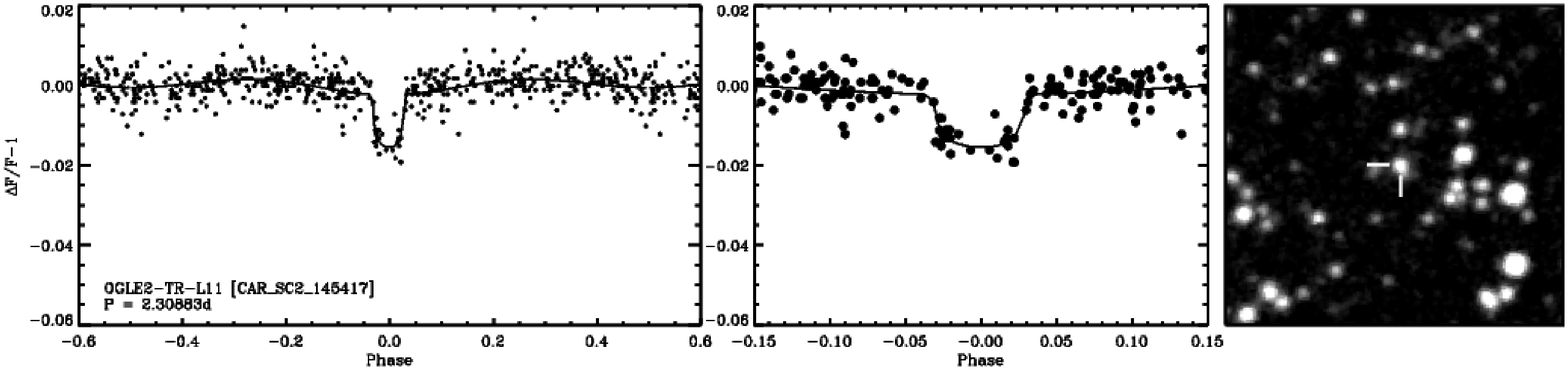,width=1.0\textwidth}
\psfig{figure=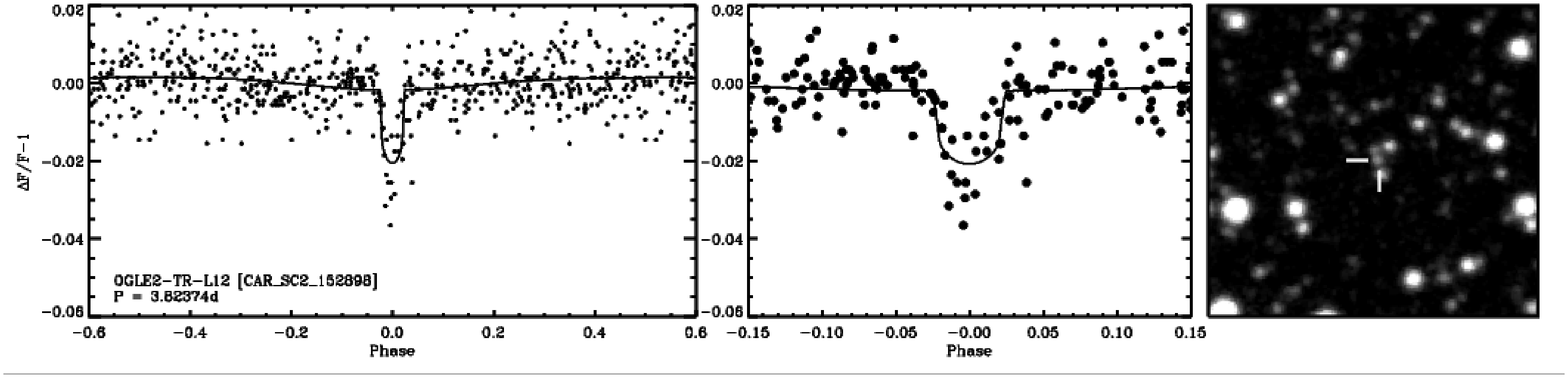,width=1.0\textwidth}
\psfig{figure=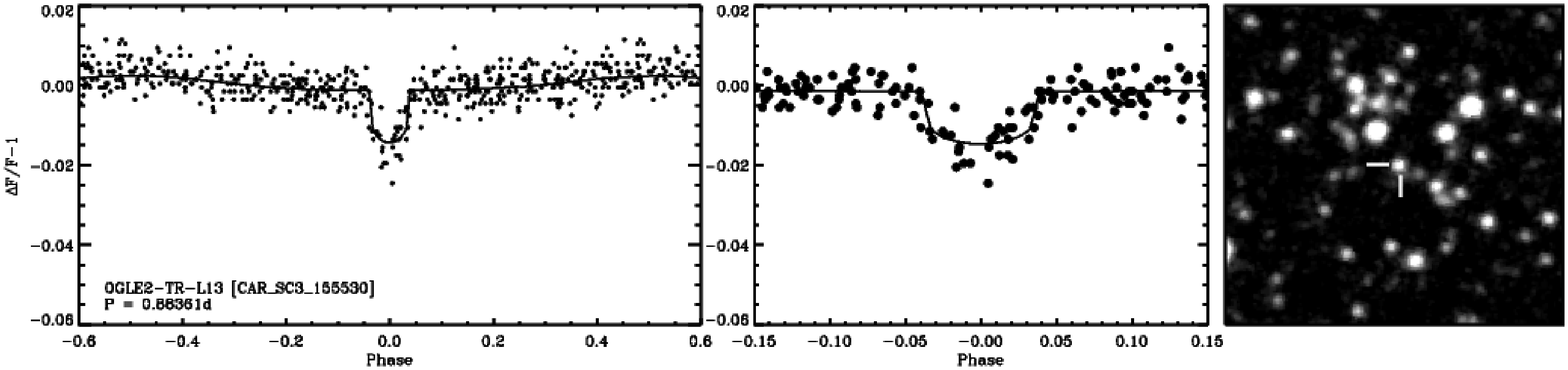,width=1.0\textwidth}
\caption{Continued....}
\end{figure*}

\end{document}